\documentclass[prl,twocolumn,showpacs]{revtex4}
\usepackage{amsmath,bm,cancel}
\usepackage{url}
\usepackage{color}
\usepackage{graphicx}
\usepackage{subfigure}

\begin{document}
%\large

\title{Universal Velocity Profile for Coherent Vortices in Two-Dimensional Turbulence}

\author{M. Chertkov$^a$, I. Kolokolov$^{a,b}$, and V. Lebedev$^{a,b}$}

\affiliation{$^a$Center for Nonlinear Studies \& Theoretical Division, LANL, Los Alamos, NM 87545, USA \\
 $^b$ Landau Institute for Theoretical Physics, Moscow, Kosygina 2, 119334, Russia}

\date{\today}

 \begin{abstract}
Two-dimensional turbulence generated in a finite box produces large-scale coherent
vortices coexisting with small-scale fluctuations. We present a rigorous theory
explaining the $\eta=1/4$ scaling in the $V\propto r^{-\eta}$ law of the velocity spatial
profile within a vortex, where $r$ is the distance from the vortex center. This scaling,
consistent with earlier numerical and laboratory measurements, is universal in its
independence of details of the small-scale injection of turbulent fluctuations and
details of the shape of the box.
 \end{abstract}

\pacs{47.27.E-, 92.60.hk}

\maketitle

The generation of large-scale motions from small-scale ones is a remarkable property of
2D turbulence. This phenomenon is a consequence of the energy transfer to large scales
\cite{67Kra,68Lei,69Bat}, realized via inverse cascade. Simulations \cite{94SY,94Bor} and
experiments \cite{00DD,05SXP} show that accumulation of energy in a large-scale coherent
structure is observed at sufficiently long times if the friction is small enough and does
not prohibit the energy cascade from reaching the system size. Recent interest in
understanding the structure of this state was sparked by experimental \cite{08XPFS,09Sha}
and numerical \cite{07CCKL} observations of large-scale coherent vortices associated with
energy condensation in forced, bounded flows. One motivation for studying 2D turbulence
comes from its structural and phenomenological similarity to quasi-geostrophic turbulence
\cite{97Les,71Cha}, such as that observed  in planetary atmospheres \cite{85NG}. Also as
suggested in \cite{05SXP}, the emergence of large-scale coherent structures in 2D is
related to the confinement transition in magnetic plasmas whose slow dynamics is
described by quasi-2D hydrodynamic equations.

In this Letter, we examine the large-scale vortices, generated by inverse energy cascade
in a finite box. We begin our discussion with a brief review of the classical theory of
inverse cascade by Kraichnan, Leith, and Batchelor (KLB) \cite{67Kra,68Lei,69Bat}. The
essential difference of 2D turbulence and 3D turbulence is the presence in the former of
a second inviscid quadratic invariant, in addition to energy, the enstrophy. Therefore,
stirring of 2D flow generates an enstrophy cascade from the forcing scale, $l$, down to
smaller scales (direct cascade) and also generates an energy cascade from the forcing
scale up scales (inverse cascade). Viscosity dissipates enstrophy at the Kolmogorov
scale, $r_\mathrm{visc}$, which is much smaller than $l$ when the Reynolds number is
large. In an infinite system, the energy cascade is eventually blocked at the scale of
$r_\mathrm{fric}$ by friction, thus resulted in establishing the two-cascade stationary
KLB turbulence for $r_\mathrm{fric}\gg l$. The Kolmogorov phenomenology (see, e.g.,
\cite{frischBook} for a review), KLB predicts the velocity power spectrum $k^{-3}$ in the
direct cascade and $k^{-5/3}$ in the inverse cascade, where $k$ is the modulus of the
wave vector. These KLB theoretical predictions were confirmed in simulations \cite{00BCV}
and laboratory experiments \cite{98PT,02KG,06CEERWX}. (Note also discussion of
experimental evidence of simultaneous inverse and forward cascade in the infinite system
setting \cite{98Rut}.) If the frictional dissipation is weak and $r_\mathrm{fric}$
exceeds the system size $L$, then ultimately the KLB regime is not applicable and a
large-scale coherent flow (occasionally called a condensate) emerges \cite{93SY}.

\vspace{-0.9cm}
\begin{figure}[ht]
\centering
      %\subfigure{\includegraphics[width=0.23\textwidth]{omegaProfilesLogLog.color.eps}}
      %\subfigure{\includegraphics[width=0.2\textwidth]{Vorticity_Profile_cond.eps}}
      \includegraphics[width=0.45\textwidth]{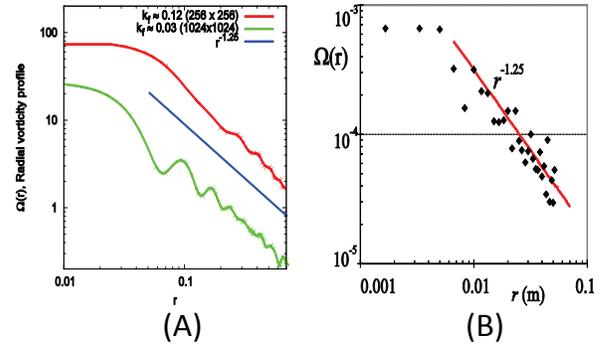}
      \vspace{-0.9cm}
  \caption{Average vorticity profile observed
  in simulations (A, Fig.~2C from \cite{07CCKL}) and experiment (B, Fig.~3 from \cite{09Sha}).}
  \label{fig:Omega}
\end{figure}
\vspace{-0.3cm}

Laboratory experiments showed that the coherent flow contains one or two
vortices, depending on the boundary conditions, and taking roughly a half of the system size \cite{09CV,09Sha}. Numerical simulations of \cite{07CCKL} reported a well-defined scaling for the average velocity profile in the interior, $V\propto r^{-\eta}$, as a function of distance, $r$, from the vortex center, and $\eta\approx 1/4$. Similar scaling was also observed in the thin layer experiment \cite{09Sha}.
Fig.~\ref{fig:Omega} summarizes the results of \cite{07CCKL,09Sha} for the
average vorticity, $\Omega$, demonstrating the $\propto r^{-5/4}$ behavior corresponding
to $\eta=1/4$. (We mention the experimental results of \cite{09Sha} to emphasize
emergence of the scaling range, possibly not reached in preceding experiments, e.g. \cite{09CV}, presumably because of a somehow higher surface friction.
%Discussion of experimental nuances,  especially in what concerns comparison of methods and results of \cite{09CV} and \cite{09Sha}, is left behind the scope of this theory paper.
)

Our main result is a rigorous derivation of the $\eta=1/4$ scaling and explanation of why
this scaling is universal. Our derivation is the first of the kind in the field of
turbulence as predicting universal scaling for a structure emerging as the result of a
nonlinear balance between the small-scale turbulence and the coherent structure generated
by turbulence. The key feature that allowed us to derive this analytical result, is the
smallness of the amplitude of the background velocity fluctuations in comparison with the
coherent part. In essence, this small parameter provides an asymptotically accurate
truncation of the generally infinite system of Hopf equations on the level of third order
correlation functions of velocity/vorticity. The $\eta=1/4$ scaling emerges from an
explicit solution of the resulting system of equations. The main contribution to the
third-order Hopf equation (for the third order correlation function) is associated with a
zero mode of the respective integro-differential operator representing the homogeneous
part of the equation. This result is substituted into the second-order Hopf equation,
thus treated as a linear inhomogeneous equation for the pair-correlation function, with
the third-order correlation function calculated on the previous step. Similarly, the
first order Hopf equation is a linear inhomogeneous equation with respect to the first
moment (the coherent term), resulting in a closed expression for the scaling exponent.
Our strategy below is to derive the set of equations, introduce the truncations and show
that a scale-invariant expression with $\eta=1/4$ gives a solution consistent with the
assumptions made in the process of the derivations.

The 2D velocity field $\bm v$ is assumed controlled by the Navier-Stocks equation
 \begin{eqnarray}
 \partial_t \omega+({\bm v}\nabla)\omega
 =-\alpha \omega +\nu\nabla^2 \omega+ \mathrm{curl}\,{\bm f},
 \label{GENNS}
 \end{eqnarray}
formulated in terms of the vorticity $\omega=\mathrm{curl}\,\bm v$. One assumes that the
fluid is incompressible, $\nabla\cdot{\bm v}=0$. The terms on the right-hand side of
Eq.~(\ref{GENNS}) represent the bottom friction, the viscosity and the turbulent forcing
respectively. The force per unit mass, $\bm f$, is assumed to be random, zero mean,
statistically homogeneous in space and time, and correlated at an intermediate scale,
$l$, called pumping scale. We study the case in which the pumping scale is much smaller
than the size of the system (of the box), $L$, and it is much larger than the Kolmogorov
(viscous) scale, $r_\mathrm{visc}$. The energy density (per unit mass) $\varepsilon$
injected by the forcing ${\bm f}$ in a unit of time per unit mass is considered constant
in space and time.

If one starts from zero velocity and turns on the pumping ${\bm f}$ at $t=0$,  in time
$\tau_d\sim l^{2/3}/\varepsilon^{1/3}$ a direct enstrophy cascade is established in the
range of scales between the pumping scale and the viscous scale. The establishment of
this direct cascade is followed by a much slower growth of the inverse energy cascade
from the pumping scale to larger scales. The energy-containing scale of the inverse
cascade grows, as $l_{inv}\sim \varepsilon^{1/2}t^{3/2}$, until the scale reaches the
system size $L$ at time $\tau_{inh}\sim(L^2/\varepsilon)^{1/3}$. After that the system
has no choice but to deviate from spatially homogeneous KLB regime, producing a
large-scale coherent flow. This picture is correct provided the bottom friction is
sufficiently weak, $\alpha\ll\epsilon^{1/3} L^{-2/3}$, as assumed in the following.
Establishing the spatial profile of the resulting average velocity at times
$t>\tau_{inh}$ is our main objective.

Once the large-scale coherent flow has emerged, the total velocity field, ${\bm v}(t,{\bm
r})$, can be decomposed into coherent and fluctuating parts, ${\bm v}(t,{\bm r})={\bm
V}(t,{\bm r})+{\bm u}(t,{\bm r})$. By definition of the coherent part, $\langle {\bm
u}\rangle=0$, where angular brackets indicate averaging over the temporal scale of the
vortex turnover time, $\sim L/V$. After a transient period, i.e. once the large-scale
flow has matured, the injected energy is mainly accumulated in the coherent component of
the velocity ${\bm V}(t,{\bm r})$, which grows slowly in time, $\propto\sqrt t$,
corresponding to the linear in time growth of the total energy. Ultimately, the average
velocity profile $\bm V$ is stabilized by the bottom friction, then the velocity
amplitude is determined by the balance between energy injection and dissipation, where
thus $V\propto 1/\sqrt{\alpha}$. Let us consider averaging Eq.~(\ref{GENNS})
 \begin{equation}
 (\partial_t+\alpha)\Omega +{\bm V}\nabla\Omega
 +\nabla\langle {\bm u}\varpi\rangle=0,
 \label{eqOmega}
 \end{equation}
where $\Omega=\langle\omega\rangle$ is the average vorticity and $\varpi=\mathrm{curl}\,
\bm u$ stands for respective fluctuations. In Eq.~(\ref{eqOmega}) we have ignored
viscosity and dropped pumping, both irrelevant at scales larger than $l$. Our
description, detailed below, is based on the assumption that the coherent flow dominates
fluctuations, $V\gg u$, possibly everywhere except for a small neighborhood of the vortex
core. (This assumption can be accurately verified via a self-consistent multi-step
procedure including an analysis of the multi-point correlation functions. A detailed
discussion of these technical details is postponed for a longer publication.)

In the periodic set-up, e.g. realized in simulations \cite{07CCKL}, a pair of vortexes
forming a dipole is formed, whereas in a bounded box set-up one observes a single vortex,
as seen in the experiment of \cite{09Sha}. Other structures, e.g. more than one pair of
vortices or stripes or combinations of stripes and vortexes, can also emerge in boxes of
special shapes, such as those with large aspect ratios or non-trivial topology (e.g.,
stripes and rings) \cite{96CS,03YMC}. In the following we will focus on an analysis of a
vortex which applies equally well to either of the two cases mentioned above. (Note also
that our approach to describing the shape of the vortex is based on analysis of
stochastic Navier-Stokes, and as such it is distinctly different from the
quasi-equilibrium 2d approach \cite{91RS,96CS,03YMC}, postulating a distribution of
Gibbs-kind controlled by the set of lagrangian multipliers associated with different
moments of vorticity.)

An emergence of the coherent vortices results in an inhomogeneous redistribution of
energy. After a vortex (or pair of vortexes) has emerged, the global profile of ${\bm
V}(t,\bm r)$ shows two regions, corresponding to a vortex exterior and an interior
respectively. In the vortex exterior the average velocity is estimated as $V\sim
\sqrt{\epsilon t}$ or $V\sim\sqrt{\epsilon/\alpha}$ (the latter corresponds to the
stationary case, in which the turbulence is stabilized by friction), while inside the
vortex the coherent part is much larger and (up to small variations we are ignoring) its
components are $V_\varphi=V(r)$ and $V_r=0$. Inspired by the results of the numerical
\cite{07CCKL} and laboratory \cite{09Sha} experiments, we assume that the spatial profile
of the coherent part in the interior of the vortex is algebraic, that is
 \begin{equation}
 V(t,\bm r)= V_0 (L/r)^\eta ,
 \label{Vradial}
 \end{equation}
where the distance $r$ is measured from the vortex center and $V_0$ estimates the
coherent part of the velocity in the vortex exterior. Eq. (\ref{Vradial}) is correct for
the $r_\mathrm{core}\ll r\ll L$ range. Here $r_\mathrm{core}$ is the size of the vortex
core. The term, $\bm V\nabla\Omega$, in Eq.~(\ref{eqOmega}) is zero due to the isotropy
of the vortex. Therefore the vortex profile is determined by a balance of the first and
third terms in Eq.~(\ref{eqOmega}). Obviously, Eq.~(\ref{eqOmega}) is not closed and one
naturally needs to consider an additional equation for the pair correlation function of
velocity/vorticity fluctuations inside the vortex.

In fact, it is convenient to derive these extra equations for the averages in two steps,
first rewriting Eq.~(\ref{GENNS})
 \begin{equation}
 (\partial_t+\alpha) u_r
 +\partial_\varphi \hat{\cal N}^{-1} \hat{\cal K} u_r
 -\partial_\varphi r^2 \hat{\cal N}^{-1}
 (\bm u \nabla \varpi) =0 \,,
 \label{ydif1}
 \end{equation}
where both the force and the viscosity terms are dropped. The differential operators in
Eq.~(\ref{ydif1}) are
 \begin{eqnarray} &&
 \hat{\cal N}
 =r[(\partial_\varrho+1)^2+\partial_\varphi^2] \,,
 \label{zdif4} \\ &&
 \hat{\cal K}=V(\partial_\varrho^2+2\partial_\varrho+2
 +\partial_\varphi^2) -(\partial_\varrho^2 V) \,,
 \label{zdif5}
 \end{eqnarray}
where $\varrho=\ln(r/L)$. Then, we introduce the pair correlation function of the radial
velocity fluctuations
 \begin{equation}
 \Phi(t,r_1,r_2,\varphi)=
 \langle u_r(t,r_1,\varphi_1)u_r(t,r_2,\varphi_2)\rangle  \,,
 \label{zdefF}
 \end{equation}
where $\varphi=\varphi_1-\varphi_2$. The correlation function is invariant under the
transformation, $\varphi\to-\varphi$, $r_1\leftrightarrow r_2$, corresponding to the
permutation of the points labeled by $1$ and $2$. Using this property and assuming
analyticity of the pair correlation function (\ref{zdefF}) for small $\varphi$ and
$\rho=\ln(r_1/r_2)$, we derive the following expression for the single-point cross-object
of the second-order, $\langle u_r \varpi \rangle$, appearing in Eq.~(\ref{eqOmega}):
 \begin{equation}
 \langle u_r \varpi \rangle=
 -\left.\frac{2}{r}\hat B\partial_\rho\partial_\varphi^{-1}
 \Phi(r,\rho,\varphi)\right|_{\rho=0,\varphi=0} \,,
 \label{zomvr}
 \end{equation}
where $r=\sqrt{r_1r_2}$ and $\hat B=1+r\partial_r/2$. Note that only antisymmetric in
$\varphi$ term in $\Phi$ contributes to $\langle u_r\varpi\rangle$.

Multiplying Eq.~(\ref{ydif1}) by the velocity at another spatial point and averaging the
resulting equation over fluctuations, one derives
 \begin{eqnarray}
 \hat{\cal N}_1^{-1}\hat{\cal N}_2^{-1}
 \left(\hat{\cal N}_2\hat{\cal K}_1
 -\hat{\cal N}_1\hat{\cal K}_2\right)
 \Phi(r_1,r_2,\varphi)
 \nonumber \\
 =r_1^2\hat{\cal N}_1^{-1} \nabla_1
 \left\langle \bm u(\bm r_1)\varpi(\bm r_1)
 u_r(\bm r_2)\right\rangle
 \nonumber \\
 -r_2^2\hat{\cal N}_2^{-1} \nabla_2
 \left\langle u_r(\bm r_1)
 \bm u(\bm r_2)\varpi(\bm r_2)\right\rangle  \,,
 \label{epair}
 \end{eqnarray}
where the irrelevant (asymptotically small) terms, containing the time derivative and the
friction coefficient $\alpha$, are omitted. The operator on the left hand side of
Eq.~(\ref{epair}) can be rewritten as
 \begin{eqnarray}
 \hat{\cal N}_2 \hat{\cal K}_1-
 \hat{\cal N}_1 \hat{\cal K}_2=
 (r_2 V_1-r_1 V_2) \times
 \nonumber \\
 \biggl\{[\partial_\rho^2+\partial_\varphi^2+\hat B^2]^2
 +(1-\eta^2)[\partial_\rho^2+\partial_\varphi^2+\hat B^2]
 \nonumber \\
 -4\hat B^2 \partial_\varrho^2
 +2\hat B(1-\eta^2)\coth[(1+\eta)\rho/2]\partial_\rho \biggr\}.
 \label{zpol5}
 \end{eqnarray}

When the separation $\bm r_1-\bm r_2$ is sufficiently small, the right hand side of
Eq.~(\ref{epair}) controls the inverse energy flux, exactly as in the traditional KLB
case. Indeed, in the spatially homogeneous case, the correlation function $\left\langle
u_\alpha(\bm r_1)\varpi(\bm r_1) u_\beta(\bm r_2)\right\rangle\sim \epsilon$ depends
solely on $\bm r_1-\bm r_2$ and it is also divergentless due to $\nabla\bm u=0$.
Substituting the Kolmogorov estimate $\Phi-\Phi(0)\sim (\epsilon|\bm r_1-\bm r_2|)^{2/3}$
into the left hand side of Eq.~(\ref{epair}), one finds that the term is negligible for
$|\bm r_1-\bm r_2|=r_{12}\ll \zeta_\star$, where $\zeta_\star(r)=\alpha^{3/4} r^{3/2}
\epsilon^{-1/4}(r/L)^{3\eta/2}$ is thus an important scale dependent on the distance from
the vortex core, $r$. One concludes that for $r_{12}\gg \zeta_\star(r)$, the inverse
cascade is modified by the coherent flow.

However, due to isotropy the term in $\Phi$ related to the KLB cascade does not
contribute to the object of our prime attention, $\langle {\bm u}\varpi\rangle$. First,
we look for such solutions, also remaining regular at small $r_{12}$, in terms of the
zero modes of the operator on the left-hand side of Eq. (\ref{epair}), thus ignoring the
smaller right-hand side in the equation. However, these zero modes do not contribute to
$\langle u_r\varpi\rangle$, because the last term in the operator (\ref{zpol5}) prohibits
odd in $\rho$ zero modes to be regular for small $\rho$. Therefore, to extract a non-zero
contribution to $\langle u_r \varpi\rangle$, one has to account for a correction to
$\Phi$, $\delta\Phi$, related to the right-hand side of Eq.~(\ref{epair}). To get a
non-trivial contribution one ought to carry our analysis to the next order in the Hopf
hierarchy describing the triple velocity correlation function, $F=\langle v_r(\bm r_1)
v_r(\bm r_2) v_r(\bm r_3)\rangle$.

The principal terms in the third-order Hopf equation governing $F$ are
 \begin{equation}
 \left\{\frac{\partial}{\partial\varphi_1}
 \hat{\cal N}_1^{-1}\hat{\cal K}_1
 \!+\!\frac{\partial}{\partial\varphi_2}
 \hat{\cal N}_2^{-1}\hat{\cal K}_2
 \!+\!\frac{\partial}{\partial\varphi_3}
 \hat{\cal N}_3^{-1}\hat{\cal K}_3\right\} F=0.
 \label{evolu}
 \end{equation}
where we again omitted asymptotically irrelevant terms, including the time derivative
term, the friction term and also the contribution related to the fourth order correlation
function. Formally, any zero mode of the operator $K$ satisfies Eq.~(\ref{evolu}), and
the quest is to find the scale-invariant zero mode of $K$, $Z_m=\exp(im\varphi
+\beta_m\varrho)$ where $m$ is an integer (to guarantee smoothness at the smallest
scales) and $\beta_m =\sqrt {m^2+\eta^2-1}-1$, which generates a nonzero contribution
into $\langle u_r \varpi\rangle$ via Eqs.~(\ref{epair},\ref{zomvr}) and has the smallest
possible $\beta_m$. The first terms in the hierarchy of possible candidates are
 \begin{equation}
 F\propto Z_m(r_1,\varphi_1) Z_{k}(r_2,\varphi_2)
 Z_{-m-k}(r_3,\varphi_3)+\dots,
 \label{striple}
 \end{equation}
where the dots represent the sum of the terms that are obtained from the first product in
Eq.~(\ref{striple}) by permuting the indices $1,2,3$. However, the expression
(\ref{striple}) generates an odd outcome for the right-hand side of Eq. (\ref{epair}),
thus resulting in an even correction $\delta\Phi$ giving no contribution to $\langle
u_r\varpi\rangle$. Therefore, one has to look for higher order terms in the hierarchy.
One finds that the desired zero mode can can be constructed with the help of an auxiliary
object, $X_m=\exp[im\varphi +(\beta_m+1+\eta)\varrho]$, satisfying $(\hat{\cal N}_m)^{-1}
\hat{\cal K}_m X_m= A_m Z_m$, where $A_m$ are real numbers:
 \begin{eqnarray}
 F=\alpha_m X_m(r_1,\varphi_1) Z_k(r_2,\varphi_2)Z_{-m-k}(r_3,\varphi_3)+\dots
 \nonumber \\
 +\alpha_k Z_m(r_1,\varphi_1) X_k(r_2,\varphi_2)Z_{-m-k}(r_3,\varphi_3)+\dots
 \nonumber \\
 +\alpha_{-k-m} Z_m(r_1,\varphi_1) Z_{k}(r_2,\varphi_2)X_{-m-k}(r_3,\varphi_3)
 +\dots,
 \label{triplec}
 \end{eqnarray}
and the dots stand for the sum of terms which accounts for respective permutations.
Eq.~(\ref{triplec}) is a solution of Eq. (\ref{evolu}), provided $\alpha_m m A_m+\alpha_k
k A_k -\alpha_{-m-k}(m+k)A_{-m-k}=0$. Choosing $m=k=1$ we find a term giving a non-odd
$\propto r^{3\eta+\sqrt{3+\eta^2}-3}$ contribution to the right hand side of Eq.
(\ref{epair}). Then the correction to the pair correlation function is $\delta\Phi\propto
r^{4\eta+\sqrt{3+\eta^2}-2}$. This result, finally substituted into the last term in Eq.
(\ref{eqOmega}), translates into the $4\eta-3+\sqrt{3+\eta^2}=-\eta$ relation, whose
solution is $\eta=1/4$. To conclude,  the $ZZZ$ and $ZZX$ terms, represented by
Eqs.~(\ref{striple}) and Eqs.~(\ref{triplec}),  are the only structures possibly
contributing to the third-order correlation function $F$, and of the various allowed
(nonzero) contributions, the $ZZX$ term (\ref{triplec}) with $m=k=1$ dominates $F$ at
$r_{1,2}\ll L$.

Substituting the expressions, corresponding to $\eta=1/4$, into the the Hopf equations of
the first, second and third orders and estimating all the terms dropped in the derivation
process confirms the validity of our asymptotic approximations. This completes our
derivations.

We now summarize our results. The main and somewhat surprising result we just derived
conserns universality of the vortex mean profile. The scaling of the vortex shape is
controlled primarily by a nontrivial zero mode of the operator on the right-hand side of
Eq.~(\ref{evolu}) and otherwise it follows from scaling relations between pairs of terms
in the first- and second-order Hopf equations. Nothing in this solution is sensitive to
the geometry of the box, or the details of the pumping. The solution also does not depend
on the type of viscosity (hyper or normal), or the damping coefficient. Our conclusion
does not depend of whether or not the coherent part grows in time or if it was already
saturated by damping. Finally, our results make predictions going far beyond the main
scaling statement, in particular detailed structure of angular harmonics is predicted for
pair and triple correlation functions in the coherent regimes. Our theoretical statements
call for accurate experimental and numerical tests.

We conclude by mentioning a number of other comprehensive questions raised by this study.
Suppose a vortex or a pair of vortexes, internally tuned and built up from the energy
flux are produced, but then the pumping is switched off. Will the initially formed vortex
will keep its shape dynamically? Also if a somewhat different in shape, non-universal and
large scale vortex is created, will it transform via decaying turbulence into the
universal shape predicted above? We conjecture that answers to both the questions are
affirmative. These questions certainly require careful investigation in the future.

The work at LANL was carried out under the auspices of the National Nuclear Security
Administration of the U.S. Department of Energy at Los Alamos National Laboratory under
Contract No. DE-AC52-06NA25396. The work of I.K. and V.L. was partially supported by RFBR
under grant no. 09-02-01346-a and FTP ``Kadry''.

\bibliographystyle{apsrev}
\bibliography{vortex}

 \end{document}